\documentclass[10pt,a4paper]{article}
\usepackage{amsmath}
\usepackage{amsfonts}
\usepackage{amssymb}
\usepackage{amsthm}
\usepackage{placeins}
\usepackage{changepage}
\usepackage{graphicx}
\usepackage{adjustbox}
\usepackage{xspace}
\usepackage{hyperref}
\usepackage{booktabs}
\usepackage{chicago}
\usepackage{url}

\newcommand{\Ebb}{\ensuremath{\mathbb{Ebb}}}

\newcommand{\Pbb}{\ensuremath{\mathbb{P}}}
\newcommand{\Fca}{\ensuremath{{\mathcal{F}}}}

\newcommand{\Qbb}{\ensuremath{{\mathbb{Q}}}}

\newcommand{\Rbb}{\ensuremath{\Xi}}

\newcommand{\CVA}{\ensuremath{\text{CVA}}}

\newcommand{\FVA}{\ensuremath{\text{FVA}}}

\newcommand{\XVA}{\ensuremath{\text{CVA+FVA}}}

\newcommand{\LGD}{\ensuremath{{L_{GD}}}}

\newcommand{\ccva}{\ensuremath{\text{CD.CVA}}}

\newcommand{\ccfa}{\ensuremath{\text{CD.FVA}}}

\newcommand{\CCVAll}{\ensuremath{\text{Climate\ Change}}\ \ensuremath{\text{Valuation\ Adjustment}}}
\newcommand{\ccvall}{\ensuremath{\text{CCVA}}}

\newcommand{\ccs}{\ensuremath{\text{Climate\ Change}}}
\newcommand{\mi}{\ensuremath{\text{Market\ Practice}}}
\newcommand{\cvaMI}{\ensuremath{\text{CVA}_\mi}}
\newcommand{\cvaCCS}{\ensuremath{\text{CVA}_\ccs}}
\newcommand{\cvami}{\ensuremath{\text{CVA}_\text{MP}}}
\newcommand{\cvaccs}{\ensuremath{\text{CVA}_\text{CC}}}
\newcommand{\fvaMI}{\ensuremath{\text{FVA}_\mi}}
\newcommand{\fvaCCS}{\ensuremath{\text{FVA}_\ccs}}
\newcommand{\fvami}{\ensuremath{\text{FVA}_\text{MP}}}
\newcommand{\fvaccs}{\ensuremath{\text{FVA}_\text{CC}}}

\newcommand{\Sparam}{\ensuremath{S(1_\text{transient},(t_\text{start}, h_\text{start});} \ensuremath{m, w; u, (t_\text{end}, h_\text{max}))}}

\theoremstyle{definition}
\newtheorem{definition}{Definition}

\begin{document}
\title{\CCVAll\ (\ccvall) 
	\\using parameterized climate change impacts}

\author{Chris Kenyon\footnote{Contact: Chris.Kenyon@mufgsecurities.com.  This paper is a personal view and does not represent the views of MUFG Securities EMEA plc (“MUSE”).  This paper is not advice.  Certain information contained in this presentation has been obtained or derived from third party sources and such information is believed to be correct and reliable but has not been independently verified.  Furthermore the information may not be current due to, among other things, changes in the financial markets or economic environment.  No obligation is accepted to update any such information contained in this presentation.  MUSE shall not be liable in any manner whatsoever for any consequences or loss (including but not limited to any direct, indirect or consequential loss, loss of profits and damages) arising from any reliance on or usage of this presentation and accepts no legal responsibility to any party who directly or indirectly receives this material.}\ , Mourad Berrahoui\footnote{Contacts: Mourad.Berrahoui@lloydsbanking.com.  The views expressed in this presentation are the personal views of the author and do not necessarily reflect the views or policies of current or previous employers. Not guaranteed fit for any purpose. Use at your own risk.}}

\date{2021-05-14, \ \ \ \ \ \ version 2.01 short\ \ \  \ to appear in {\bf Risk}}

\maketitle
	
\begin{abstract}
	We introduce \CCVAll\ to capture climate change impacts on \XVA\ that are currently invisible assuming typical market practice.  
\end{abstract}
	
\tableofcontents

\section{Introduction}

Climate change risk comprises physical, transition, and liability risks to assets, companies and sovereign entities\footnote{https://www.bankofengland.co.uk/knowledgebank/climate-change-what-are-the-risks-to-financial-stability} \cite{boe2019ss3,ecb2020guide}.  
Credit valuation adjustment (CVA) quantifies expected loss on counterparty default \cite{green2015xva,basel2021cva}, and the costs of funding are captured in funding valuation adjustment (FVA), together \XVA.  However, CVA and FVA are based on extrapolation of credit default swap (CDS) spreads which are typically traded only up to 10 year maturity, see Table \ref{t:dtcc}, and inclusion of bond trading where applicable.

We introduce \CCVAll\ (CCVA) to capture the difference in expected loss and funding  between usual credit information extrapolation and the parameterized inclusion of economic stress from climate change.  The parameterization we introduce flexibly captures both climate endpoints, and transition effects.  Climate endpoints, like sea level rise, can have significant effects on \XVA\ even if the  climate endpoint is at the end of the century and trades in scope of \XVA\ have 20 to 30 year maturity.   We also provide a quantification of the relationship between transition effects and \XVA\ impacts for example trades.
 
Climate change valuation adjustment will be negative in cases where climate change has favorable outcomes, i.e. lower cost.  Examples may include technology providers with long development cycles that address climate mitigation, and regions where there are beneficial effects.
 
The contributions of this paper are: firstly the introduction of \CCVAll\ to capture climate change impacts on \XVA\ that are currently invisible assuming typical market practice; secondly the introduction of a flexible, and expressive instantaneous hazard rate parametrization to capture the path to climate change endpoints, and for transition effects; and thirdly a quantification of examples of typical interest where there is risk of economic stress from sea level change, or change in business model.

\section{Context}

We will define the \ccvall\ as climate related expected loss and funding costs that are not already captured by \CVA\ and \FVA\ calculated by usual market practice.  Thus \ccvall\ captures the difference between 1) combined market implied and physical measure expected loss considering the economic impact of climate change, and 2) typical bank market implied expected loss from some extrapolation of hazard rates outside the maturity of liquid CDS.    In order to make this definition precise we must first describe and define the concepts and probability spaces involved.  Before this we recall limitations of the CDS market.

\subsection{Data limitations driving market practice of \XVA}

Market implied counterparty default probability is inferred from spreads of traded CDS, augmented by bonds where applicable.    However, few CDS are traded beyond 5 years and almost none beyond 10 years.  Many counterparties, e.g. project finance, have no CDS and so are priced and hedged primarily from CDS proxies.  For proxy curves, CDS indices can be particularly important.   Table \ref{t:dtcc} shows volumes for CDS indices from a Swaps Data Repository (DTCC).  CDS indices are more traded than single name but few are even defined beyond 10 years: we see 98\%\ of reported trading volume on DTCC is for maturities up to 5 years.  

Given the lack of actual transactions, market practice is to use some form of extrapolation beyond 10 years.  Ratings may inform bond prices and proxy CDS curves, but corporate ratings typically have only a three to five year look ahead \cite{fitch2020corporate}.  

\begin{table}[h]
\begin{adjustwidth}{-3cm}{-3cm}
\begin{center}
\begin{adjustbox}{max width=1.5\textwidth}
\begin{tabular}{lrrrrrrrrrrr}
\textbf{Cumulative notional by index}   & \multicolumn{11}{c}{\textbf{CDS maturity rounded to neared year}}                                                                                                                               \\
\textbf{on DTCC 2021-01-21 to   2021-02-19} & \textbf{0}     & \textbf{1}     & \textbf{2}     & \textbf{3}     & \textbf{4}     & \textbf{5}      & \textbf{6}      & \textbf{7}      & \textbf{8}      & \textbf{9}      & \textbf{10}      \\
CDX:CDXEmergingMarkets         & 0\%            & 0\%            & 0\%            & 0\%            & 7\%            & 100\%           &                 &                 &                 &                 &                  \\
CDX:CDXHY                      & 0\%            & 0\%            & 1\%            & 2\%            & 5\%            & 100\%           &                 &                 &                 &                 &                  \\
CDX:CDXIG                      & 0\%            & 0\%            & 1\%            & 3\%            & 9\%            & 98\%            & 98\%            & 98\%            & 98\%            & 99\%            & 100\%            \\
iTraxx:iTraxxAsiaExJapan       & 0\%            & 0\%            & 0\%            & 0\%            & 9\%            & 100\%           &                 &                 &                 &                 &                  \\
iTraxx:iTraxxAustralia         & 0\%            & 0\%            & 0\%            & 0\%            & 20\%           & 100\%           &                 &                 &                 &                 &                  \\
iTraxx:iTraxxEurope            & 0\%            & 1\%            & 3\%            & 7\%            & 10\%           & 98\%            & 98\%            & 98\%            & 99\%            & 99\%            & 100\%            \\
iTraxx:iTraxxJapan             & 0\%            & 0\%            & 0\%            & 0\%            & 0\%            & 100\%           &                 &                 &                 &                 &                  \\
\textbf{Grand Total}                        & \textbf{0.0\%} & \textbf{0.5\%} & \textbf{1.8\%} & \textbf{3.9\%} & \textbf{8.5\%} & \textbf{98.5\%} & \textbf{98.6\%} & \textbf{98.8\%} & \textbf{98.8\%} & \textbf{99.0\%} & \textbf{100.0\%}
\end{tabular}
\end{adjustbox}
	\caption{Cumulative CDS transaction volume for indices referring to corporates on DTCC over a recent 30-day period, 2021-01-19 to 2021-02-20.   DTCC is a US Swaps Data Repository so sees mostly US transaction. CDS indices are more traded than single-name CDS.}
	\label{t:dtcc}
\end{center}
\end{adjustwidth}
\end{table}

\subsection{Source of \CCVAll}

\cvaMI\ does not incorporate climate related risk where this has effect beyond 5 or 10 years because of how \cvaMI\ is calculated.  \cvaMI\ is based on CDS data.  CDS data is market-based up to 5 or 10 years and then typically extrapolated flat judging by data from CDS runs (i.e. strips of tradeable CDS quotes) and CDS transaction repositories.  Note that CDS data providers typically provide extrapolated CDS values using their own internal models  beyond 10 years when they have insufficient contributors, obviously this is not tradable data.    We capture the climate change difference of \cvaCCS\  versus market CDS with flat extrapolation using \ccva, and the similar effect on \FVA\ by \ccfa\ defined below in Sections \ref{s:prob} and \ref{s:def}.

\cvaMI\ is priced using a market-implied methodology  but it is not hedged in practice beyond 10 years judging from transaction repository data, e.g. from DTCC\footnote{\url{https://www.dtcc.com/}}.  Thus banks face climate change, and other, risks on derivatives over 10 years because banks do not hedge these risks in the CDS market in as much as transaction repositories are reflective of trading. 

For entities that face no climate change risk \ccvall\ will be zero.  For entities where market practice already incorporates climate change risk \ccvall\ will also be zero.  Considering CDS market transactions, and market practice detailed above, no information beyond 5 or 10 years is incorporated into \CVA\ nor \FVA\ hence \ccvall\ will be non-zero for trades with entities that face climate related risks or benefits outside the 5 or 10 year horizon of market traded CDS data.  Where counterparties are on proxy CDS curves \ccvall\ can be non-zero for any length contract where counterparty-specific climate risk is not included in the proxy.

\subsection{Market-implied measure and physical measures}  \label{s:prob}

 Market data may define a unique market implied measure \Qbb, but physical measures \Pbb\ are always subjective as they derive from a choice of calibration.

  Since \ccvall\ is based on model predictions rather than tradable instruments it is a \Pbb-measure quantity.  Standard CVA may be thought of as a \Qbb\ measure quantity.  However, because of the lack of hedging beyond 5 to 10 years it is a mix between replication-based pricing and a measure represented by the CDS extrapolation.  We shall label this measure given by market practice of CDS extrapolation $\Xi$ (Xi for eXtrapolation).

We want to be able to price \CVA\ and \FVA\ as banks normally price them and to price \ccvall.  For normal bank pricing we introduce the probability space:
\[
X = (\Omega,\Fca,\Pbb)
\]
on a set of events $\Omega(t)$ with a filtration $\Fca(t)$ and corresponding probability measures $\Pbb(t)$.  The equivalent probability space with a risk-neutral measure, given that the last traded CDS maturity is $T$, is
\[
Y_{Q\Xi}(T) = (\Omega,\Fca, [\Qbb; T; \Rbb])
\]
on events $\Omega_{\le T} = \Omega(t) s.t.\ t \le T$ with filtration $\Fca_{\le T} = \Fca(t)\ s.t.\ t \le T$ and risk neutral measure \Qbb\ on $\Fca_T$.  Note that the risk neutral measure only exists for $t \le T$.
We use the measure \Rbb\ for $t>T$ on events $\Omega_{>T} = \Omega(t)\ s.t.\ t > T$ with filtration $\Fca_{>T} = \Fca(t)\ s.t.\ t > T$.  \Rbb\ is defined as a measure in which non-credit items can be hedged but credit items cannot be hedged but are priced assuming that CDS's are extrapolated flat (or according to some internal choice).  We assume independence of credit and non-credit events for simplicity.  

Note that \Rbb\ is not \Pbb,  for $t > T$.  \Rbb\ can  be thought of as an extrapolation of \Qbb\ following the rule that CDS quotes are extrapolated flat, or according to a Bank's internal methodology.

To capture what may actually happen we introduce the probability space combining the risk neutral measure before $T$ and the physical measure after $T$:
\[
Y_{QP}(T) = (\Omega,\Fca, [\Qbb; T; \Pbb])
\]

\section{\CCVAll}  \label{s:def}

Now we have appropriate probability spaces and measures, we can define valuation adjustments based on market practice, and based on including climate change, and then \ccvall\ as the difference.

We define \CVA\ and \FVA\ including the measure involved, based on \cite{burgard2013funding} and then specialize these to define \ccvall.
\begin{definition}[\CVA\ and \FVA\ under probability space ${Y(\Omega,\Fca,\Gamma)}$]
\begin{align} 
\CVA^{Y(\Omega,\Fca,\Gamma)} &= \Ebb^\Gamma\left[ \int_{u=0}^{u=T}\LGD(u) \lambda(u) e^{\int_{s=t_0}^{s=u}  -\lambda(s) ds}     D_{r_F}(u)    (\Pi(u) - X(u))^+   du   \right]     \label{e:cva}\\ 
\FVA^{Y(\Omega,\Fca,\Gamma)} &=  \Ebb^\Gamma \left[ \int_{u=0}^{u=T}s_F(t) e^{\int_{s=t_0}^{s=u}  -\lambda(u) ds}    D_{r_F}(u)    (\Pi(u) -  X(u))    du  \right]     \label{e:fva}
\end{align}
where $\Pi(u)$ is the value of the portfolio with the counterparty and $X(u)$ the collateral value.
\end{definition}
\noindent
The usual market implied \CVA\ and \FVA\ based on market practice are:
\begin{definition}[Market implied \CVA\ and \FVA, \cvami\ and \fvami]
	\begin{align} 
	\cvami = \cvaMI = \CVA^{Y_{Q\Xi}} &=  \CVA^{Y(\Omega,\Fca,[\Qbb; T; \Rbb])} \label{e:cvami}\\ 
	\fvami = \fvaMI = \FVA^{Y_{Q\Xi}} &=  \FVA^{Y(\Omega,\Fca,[\Qbb; T; \Rbb])}  \label{e:fvami}
	\end{align}
\end{definition}
\noindent
\CVA\ and \FVA\ including climate change are defined similarly based on probability space used.
\begin{definition}[\CVA\ and \FVA\ including climate change, \cvaccs\ and \fvaccs]
	\begin{align} 
	\cvaccs = \cvaCCS = \CVA^{Y_{QP}} &=  \CVA^{Y(\Omega,\Fca,[\Qbb; T; \Pbb])} \label{e:cvaccs}\\ 
	\fvaccs = \fvaCCS = \FVA^{Y_{QP}} &=  \FVA^{Y(\Omega,\Fca,[\Qbb; T; \Pbb])}  \label{e:fvaccs}
	\end{align}
\end{definition}
Now we can define \ccva\ and \ccfa\ as the difference between the versions including climate change and market practice (e.g. flat CDS extrapolation).  The sum of the differences is the \ccvall.
\begin{definition}[Climate Change Valuation Adjustment, \ccvall, and climate change differences in valuation adjustments for credit and funding]
	\begin{align} 
	\ccvall &= \ccva + \ccfa  \label{e:ccvall}\\
	\ccva &= \cvaCCS - \cvaMI = \CVA^{Y_{QP}} -\CVA^{Y_{Q\Xi}}  \label{e:ccva}\\ 
	\ccfa &= \fvaCCS - \fvaMI = \FVA^{Y_{QP}} -\FVA^{Y_{Q\Xi}}  \label{e:ccfa}
	\end{align}
\end{definition}
These definitions capture what is not in the market practice valuation adjustments.  If market practice changes so that climate change is included then, e.g.  $\cvaCCS = \cvaMI$, and the differences will be zero.  Here we highlight was is not currently included.  Below we estimate the size of \ccvall\ for a particular subset of entities and situations.

Note that \ccvall\ will be less than zero for cases where climate change has beneficial effects for the counterparty.

\section{Climate economic effect parameterization}   \label{s:param}

To be able to discuss and compare paths of economic stress to climate endpoints we introduce a sigmoid parameterization of the instantaneous hazard rate, $\lambda(t)$: 
\[
\lambda(t) = \Sparam
\]
For parameter details see Tables \ref{t:sigmoid} and \ref{f:transient}.  This parameterization is expressive enough to cover different paths of economic stress buildup.  The parameterization flexibly connects the longest traded CDS maturity and level, with the climate change endpoint, by allowing specification of the mid point $m$ of the stress and the width $w$ of the stress buildup.  If we specify that instead of ending at a high hazard level the curve returns to the original level, i.e. $1_\text{transient}$ is true, then the same parameterization models transient transition effects. 

In this way we capture approach to default and transition with a single set of parameters.  These parameters can be specified for each counterparty of a bank for example by  internal credit risk management, or a regulatory body for all banks, to define climate change scenarios independent of the details of the driving mechanisms.  

Custom $\lambda(t)$ models can be specified by credit departments using integrated assessment models (IAMs).  IAM take climate scenarios, micro- and macro-economic transimission channels, and deliver economic impact scenarios \cite{nordhaus2017integrated}.  Typically there might be a two-step approach of an initial assessment by a credit department working with the relationship manager to produce sigmoid parameters that are then refined by reference to IAM scenarios based on \cite{ngfs2020guide}, or vice versa.

The hazard curve for a counterparty including climate effects consists of two parts
\begin{itemize}
\item Hedgeable section with \Qbb\ measure $\lambda(t)$ from traded CDS levels, up to 5 or 10 years
\item Sigmoid section in \Pbb\ measure, where $\lambda(t) = \Sparam$
\end{itemize}

\subsection{Sigmoid parameterization, stressed endpoint}   \label{s:sigmoid}

\begin{figure}[h]
	\begin{center}
		\includegraphics[trim=0 0 0 0, clip, width=0.75\textwidth]{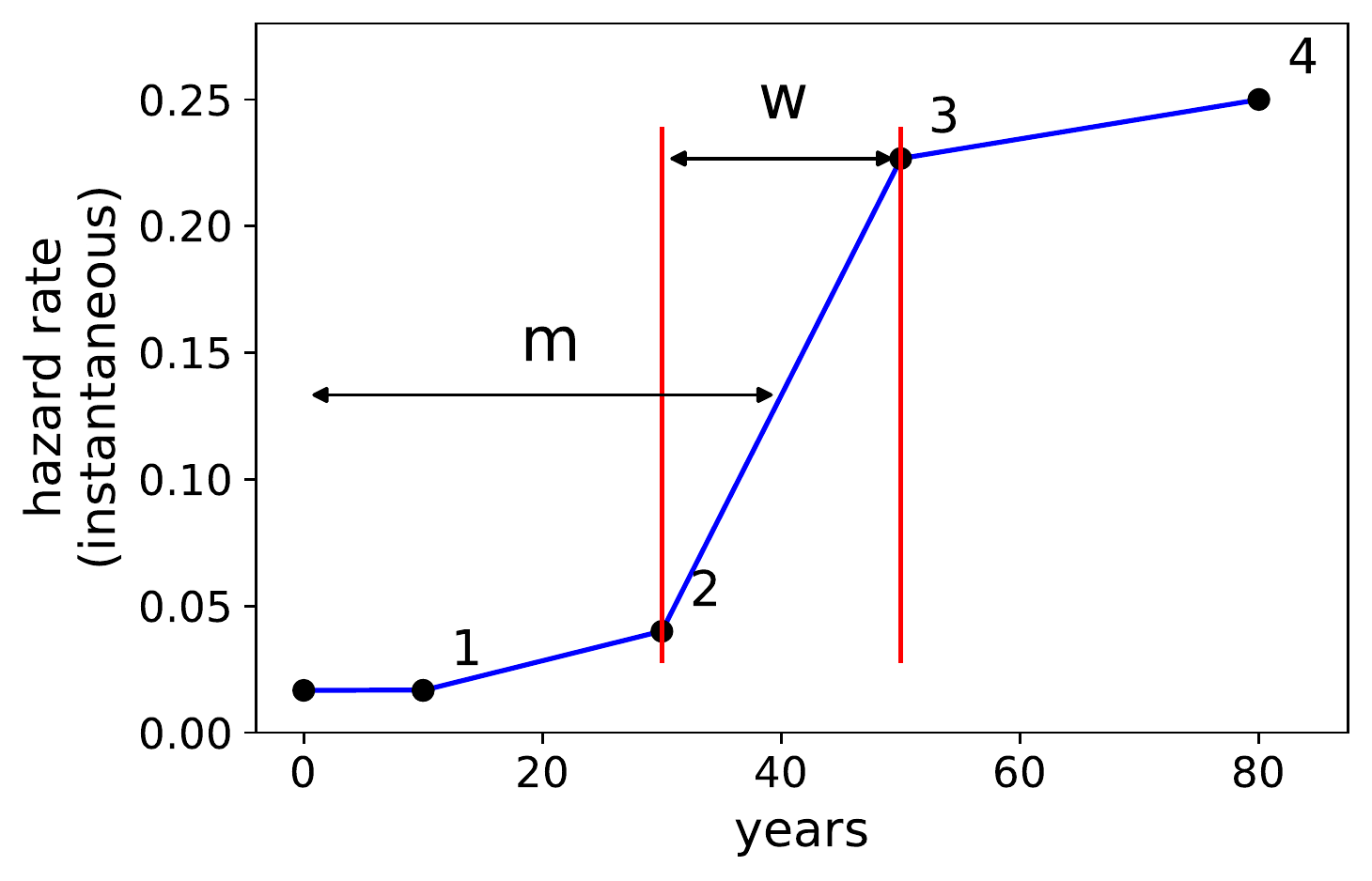}
	\end{center}
	\caption{Sigmoid parameterization for the approach of instantaneous hazard rates to default, \Sparam, with $1_\text{transient}$  False.  See Table \ref{t:sigmoid} for details.}
	\label{f:sigmoid}
\end{figure}

This sigmoid parameterization is shown in Figure \ref{f:sigmoid} with parameters described in Table \ref{t:sigmoid}.  The resulting curve is \Sparam, with $1_\text{transient}$ True. 

Note that if the slope of the last section is greater than the slope of the mid section, then point 3 is removed so there is a straight line between point 2 and point 4.  This is because it is physically reasonable to have a jump in instantaneous hazard rates in the transition from the \Qbb\ section to the \Pbb\ section, but there is no justification for a jump at the end of the \Pbb\ section. 

CDS spreads and survival probabilities are given in Table \ref{t:examples}.

\begin{table}[h]
	\begin{center}
	\begin{tabular}[htb]{rcp{6.2cm}}
		\bf Parameter & \bf Example value & \bf Description \\ \hline
		$1_\text{transient}$       & False \\
		$m$ & 40 years & time to mid-impact \\
		$w$ & 20 years & width of middle section \\
		$(t_\text{start}, h_\text{start})$ & (10, 0170)  &coordinates of end of \Qbb\ measure section and start of \Pbb\ measure section that approaches default\\
		$(t_\text{end}, h_\text{end})$ & (80, 0.2500) & coordinates of end of impact\\
		$u$ & 10\%\ & fraction of impact ($h_\text{end} - h_\text{start}$) for initial increase, and final approach to $h_\text{max}$ \\ \hline
		\bf Point & \multicolumn{2}{l}{\bf Definition} \\ \hline
		\bf 1 &  \multicolumn{2}{l}{  $(t_\text{start}, h_\text{start})$ } \\
		\bf 2 &  \multicolumn{2}{l}{  $(m - w/2, \ h_\text{start} + u \times(h_\text{max} - h_\text{start})  )$  } \\
		\bf 3 &  \multicolumn{2}{l}{  $(m + w/2, \ h_\text{max} - u \times(h_\text{max} - h_\text{start})  )$ }\ if   $1_\text{transient}= \text{False}$\\
		 &  \multicolumn{2}{l}{  $(m + w/2, \ h_\text{start} + u \times(h_\text{max} - h_\text{start})  )$ }\ if   $1_\text{transient}= \text{True}$\\
		\bf 4 &  \multicolumn{2}{l}{   $(t_\text{end}, h_\text{max})$ } \ if   $1_\text{transient}= \text{False}$\\ 
		  &  \multicolumn{2}{l}{   $(t_\text{end}, h_\text{start})$ } \ if   $1_\text{transient}= \text{True}$\\
		\bf 5 &  \multicolumn{2}{l}{   $(m, h_\text{max})$ } \ only present if   $1_\text{transient}= \text{True}$\\  \hline
	\end{tabular}
	\end{center}
	\caption{Sigmoid parameterization, and point definition, for the approach of instantaneous hazard rates to default, \Sparam.  Note that if the slope of the last section is greater than the slope of the mid section, then point 2 is removed so there is a straight line between point 2 and point 4.  See Figure \ref{f:sigmoid} for graphical view using the example parameters.}
	\label{t:sigmoid}
\end{table}

\subsection{Sigmoid parameterization, transient transition effects}   \label{s:transition}

\begin{figure}[h]
	\begin{center}
		\includegraphics[trim=0 0 0 0, clip, width=0.75\textwidth]{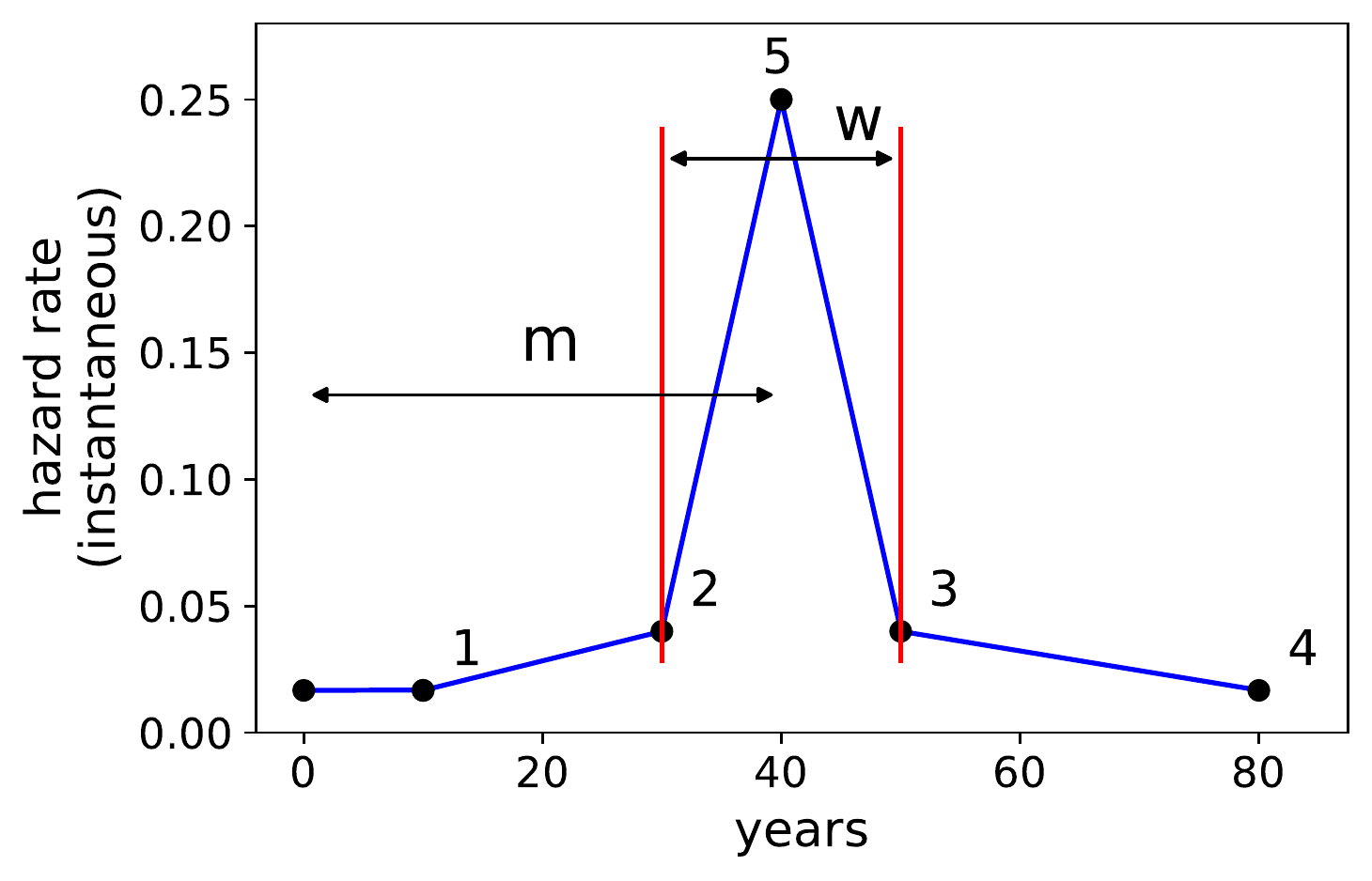}
	\end{center}
	\caption{Sigmoid parameterization for modeling of transition stress uses the same parameters, \Sparam, but now with $1_\text{transient}$ True.  See Table \ref{t:sigmoid} and Section \ref{s:transition} for details.}
	\label{f:transient}
\end{figure}

The sigmoid parameterization for a transient transition effect where economic stress returns to normal is shown in Figure \ref{f:transient}.   Parameters described in Table \ref{t:sigmoid}, except that $1_\text{transient}$ is now True.  The resulting curve is \Sparam.  Figure  \ref{f:transient} also defines the parameters.

CDS spreads and survival probabilities are also given in Table \ref{t:examples}.

\begin{table}
\centering
\begin{tabular}{rrr|rr|r}
\toprule
 maturity &  \multicolumn{2}{c}{$1_\text{transient}=$False}    &  \multicolumn{2}{c}{$1_\text{transient}=$True} &  flat CDS, 100bps \\
 years     &       CDS (bps)                                    &      survival        &       CDS (bps)                                   &    survival         &  survival\\
\midrule
        10 &                      100 &                     84.65 &                        100 &                       84.65 &                  84.65 \\
        20 &                      114 &                     67.55 &                        114 &                       67.55 &                  71.65 \\
        30 &                      132 &                     47.97 &                        132 &                       47.97 &                  60.65 \\
        40 &                      163 &                     20.07 &                        178 &                       11.14 &                  51.34 \\
        50 &                      180 &                      3.30 &                        188 &                        2.64 &                  43.46 \\
        60 &                      183 &                      0.33 &                        188 &                        1.84 &                  36.79 \\
        70 &                      183 &                      0.03 &                        188 &                        1.39 &                  31.14 \\
        80 &                      183 &                      0.00 &                        188 &                        1.13 &                  26.36 \\
\bottomrule
\end{tabular}
\caption{CDS and survival probabilities for the two examples in Figures \ref{f:sigmoid} and \ref{f:transient}.}
\label{t:examples}
\end{table}

\FloatBarrier

\FloatBarrier

\section{Numerical examples}

We consider climate change  test cases using the sigmoid parameterization quantifing effects on at the money (ATM) USD interest rate swaps (IRS) for two sets of cases:
\begin{itemize}
\item First set of cases:   the entity has reasonable expectation of default from continually increasing economic stress caused by rising sea level.  Examples of such entities include low-lying coastal cities, and associated special purpose vehicles (SPVs) used for essential infrastructure, such as roads, bridges, tunnels, housing, etc.   
\item Second set of cases:  transient transition risks where the mid point economic stress of the transition occurs from 15 to 75 years in the future and has a duration of 1 to 10 years.  We do not need to consider transition stresses within 10 years because we assume that single name CDS are traded to 10 years and that the Bank can fully hedge \XVA\ up to 10 years.
\end{itemize}
with setup :
\begin{itemize}	
	\item asof date 2020-01-29 for USD yield curve, single curve approach.  Normal volatility, flat at 20bps.
	\item uncollateralized trade.  This is typical for infrastructure projects via SPVs.
	\item maximum instantaneous hazard rate at climate change endpoint: 2500 basis points (bps). 
	\item recovery rate on CDS, 40\%. 
	\item IRS length: 20 to 50 years.  
	\item Funding spread is 100bps, flat
	\item We assume traded CDS out to 10 years, flat, at 100 bps.
\end{itemize}

\subsection{Slowest approach to endpoint at 2051 to 2101}

Here we consider \ccvall\ for the slowest possible approach to a default instantaneous hazard rate that is reached by 2050 to 2100.  We first consider the most benign example where the climate change endpoint is reached in 80 years, and then a range of endpoint dates.

\subsubsection{Endpoint reached in 80 years}

\begin{figure}[h]
\begin{adjustwidth}{-2cm}{-2cm}
	\begin{center}
		\includegraphics[trim=0 0 0 0, clip, width=0.65\textwidth]{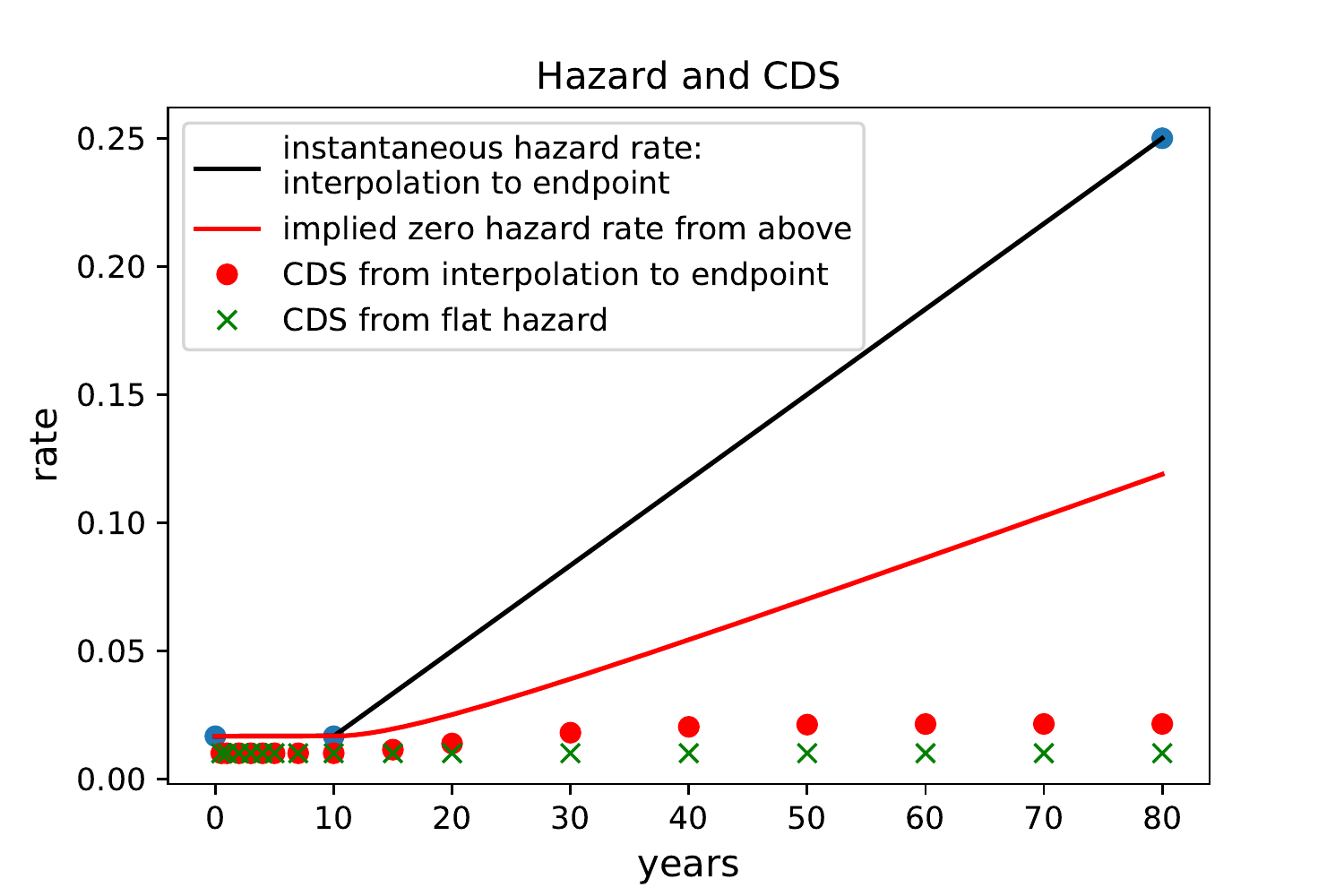}
		\includegraphics[trim=0 0 0 0, clip, width=0.65\textwidth]{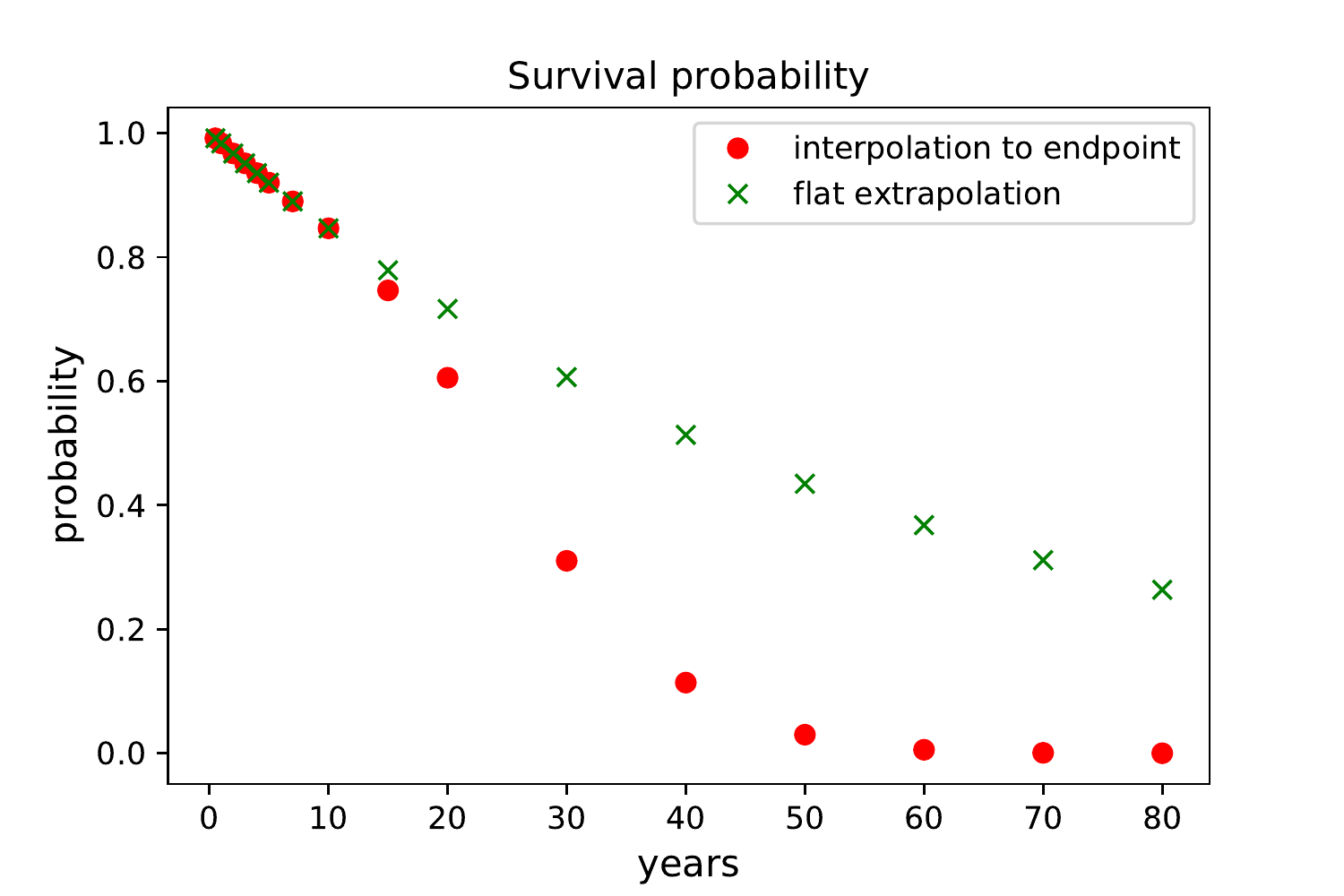}
	\end{center}
	\caption{Slowest uniform approach of instantaneous hazard rate to climate change endpoint in 80 years, starting from CDS of 100bps up to 10 years  on LEFT above, and derived zero (average) hazard rate.  On RIGHT the derived survival probabilities.}
	\label{f:ex80}
\end{adjustwidth}
\end{figure}

Figure \ref{f:ex80} shows an example of slowest uniform approach of instantaneous hazard rate to climate change endpoint in 80 years starting from CDS of 100bps up to 10 years, and the derived average hazard rates, and survival probabilities.  The derived CDS rates are shown in Table \ref{t:ex80}.  Note that we have ignored IMM dates as these have little effect on results.

We see from Figure \ref{f:ex80} and Table \ref{t:ex80} that even in one of the most benign examples we can create, i.e. start from 100 bps up to 10Y, approach climate change endpoint in 80Y, there are significant consequences for survival probabilities at 20Y and by 50Y the survival probability has almost reached zero.  In as much as there are earlier economic consequences adapting to distant (80Y) future climate endpoints can have significant earlier effects.  

Although the CDS spreads only double at 40Y to 80Y, this is deceptive.  The reason that the CDS spreads do not increase further is that both the fee and protection legs effectively cease to exist around 50Y, so further quotes carry no information.

\begin{table}
	\begin{adjustwidth}{-2cm}{-2cm}
		\begin{center}
\begin{tabular}{rrrr}
	\toprule
 maturity &  CDS(linear hazard) (bps) &  survival(linear hazard) &  survival(flat hazard) \\
\midrule
10 &                     100 &                    84.65 &                  84.65 \\
20 &                     138 &                    60.55 &                  71.65 \\
30 &                     180 &                    31.04 &                  60.65 \\
40 &                     203 &                    11.40 &                  51.34 \\
50 &                     212 &                     3.00 &                  43.46 \\
60 &                     214 &                     0.57 &                  36.79 \\
70 &                     215 &                     0.08 &                  31.14 \\
80 &                     215 &                     0.01 &                  26.36 \\
	\bottomrule
\end{tabular}
\end{center}
\caption{CDS rates implied from slowest uniform approach of instantaneous hazard rate to climate change endpoint in 80 years, starting from CDS of 100 bps up to 10 years. shown in Figure \ref{f:ex80}.  The flat CDS extrapolation is 100bps for all times.  Survival probabilities are to the maturity in the first column.}
\label{t:ex80}
\end{adjustwidth}
\end{table}

\subsubsection{Climate change endpoint reached in 30 to 80 years} \label{s:slow}

Here we give the \XVA\ changes considering climate change endpoints at 30 to 80 years against IRS of 20 to 50 year maturities.  Here the instantaneous hazard rates increase at the slowest uniform rate, i.e. a straight line from the end of the traded CDS at 10 years to the climate change endpoint.  Hazard rates are kept constant once reaching the maximum level of 2500bps.

We observe in Table \ref{t:ex3080} that there are significant effects on the CVA for all IRS, even as short as 20 years given a climate change endpoint in 2101 (i.e. 80 years from 2021), of an increase of 23\%.  The decrease in FVA, because funding costs are paid for less time partly mitigates this increase, and the overall effects is roughly a 10\%\ increase in CVA+FVA, i.e. CCVA is roughly 10\% of the value ignoring climate change.  This is the most benign case.

\begin{table}
\begin{adjustwidth}{-0cm}{-0cm}
\begin{center}
\begin{tabular}{lrrrr}
	\toprule
	{} & \multicolumn{4}{l}{change in CVA \%} \\
	IRS length (years) &         20 &     30 &     40 &     50 \\\cline{1-1}
	width (years) &            &        &        &        \\
	\midrule
	20    &       71 &  141 &  140 &  130 \\
	30    &       51 &  113 &  117 &  113 \\
	40    &       39 &   93 &  100 &  100 \\
	50    &       32 &   80 &   88 &   90 \\
	60    &       27 &   69 &   78 &   81 \\
	70    &       23 &   61 &   70 &   74 \\
	\bottomrule
\end{tabular}

\vskip5mm

\begin{tabular}{lrrrr}
	\toprule
	{} & \multicolumn{4}{l}{change in FVA \%} \\
	IRS length (years)&         20 &    30 &    40 &    50 \\\cline{1-1}
	width (years)&            &       &       &       \\
	\midrule
	20    &       -4 & -18 & -19 & -21 \\
	30    &       -3 & -13 & -15 & -16 \\
	40    &       -2 & -11 & -12 & -14 \\
	50    &       -2 &  -9 & -10 & -12 \\
	60    &       -1 &  -8 &  -9 & -10 \\
	70    &       -1 &  -7 &  -8 &  -9 \\
	\bottomrule
\end{tabular}

\vskip 5mm

\begin{tabular}{lrrrrrr}
	\toprule
	{} &  & extrapolation of &\multicolumn{4}{l}{change in \XVA\ \%} \\
	IRS length (years)&CDS slope&CDS level after &        20 &    30 &    40 &    50 \\\cline{1-1}
	width (years)& bps/year &  80 years (bps) &         &       &       &       \\
	\midrule
	20    &  125 & 8333 &    37 &  67 &  73 &  73 \\
	30    & 83 & 5611 &   26 &  54 &  62 &  64 \\
	40    & 63 &4250 &     20 &  45 &  53 &  57 \\
	50    & 50 &3433 &     17 &  39 &  47 &  51 \\
	60    & 42 &2889 &     14 &  34 &  42 &  47 \\
	70    & 36 &2500 &     12 &  30 &  38 &  43 \\
	\bottomrule
\end{tabular}

\end{center}
\caption{Slowest uniform increase in hazard rate results.  Changes in CVA (top), FVA (mid), and CVA+FVA (bottom), i.e. relative sizes of CD.CVA, CD.FVA, and CCVA compared to flat CDS extrapolation.  Notice that increased hazard rates is beneficial for FVA but not so for CVA. FVA and CVA are different sizes so the overall result is not a simple average.}
\label{t:ex3080}
\end{adjustwidth}
\end{table}

\subsection{Impact around midpoint to 2101}   \label{s:mid}

Here we assume that the impact on the instantaneous hazard rate is around the mid point of the time to the climate change endpoint.  We also assume that there is a 5\%\ build-up, and approach to maximum instantaneous hazard rate, i.e. $u=0.05$.

Results are shown in Table \ref{t:exmid}.  We see that the effects are much milder than with a uniform build up of economic stress, essentially because we are assuming a delay on the economic impact.

\begin{figure}[h]
	\begin{adjustwidth}{-2cm}{-2cm}
		\begin{center}
			\includegraphics[trim=0 0 0 0, clip, width=0.65\textwidth]{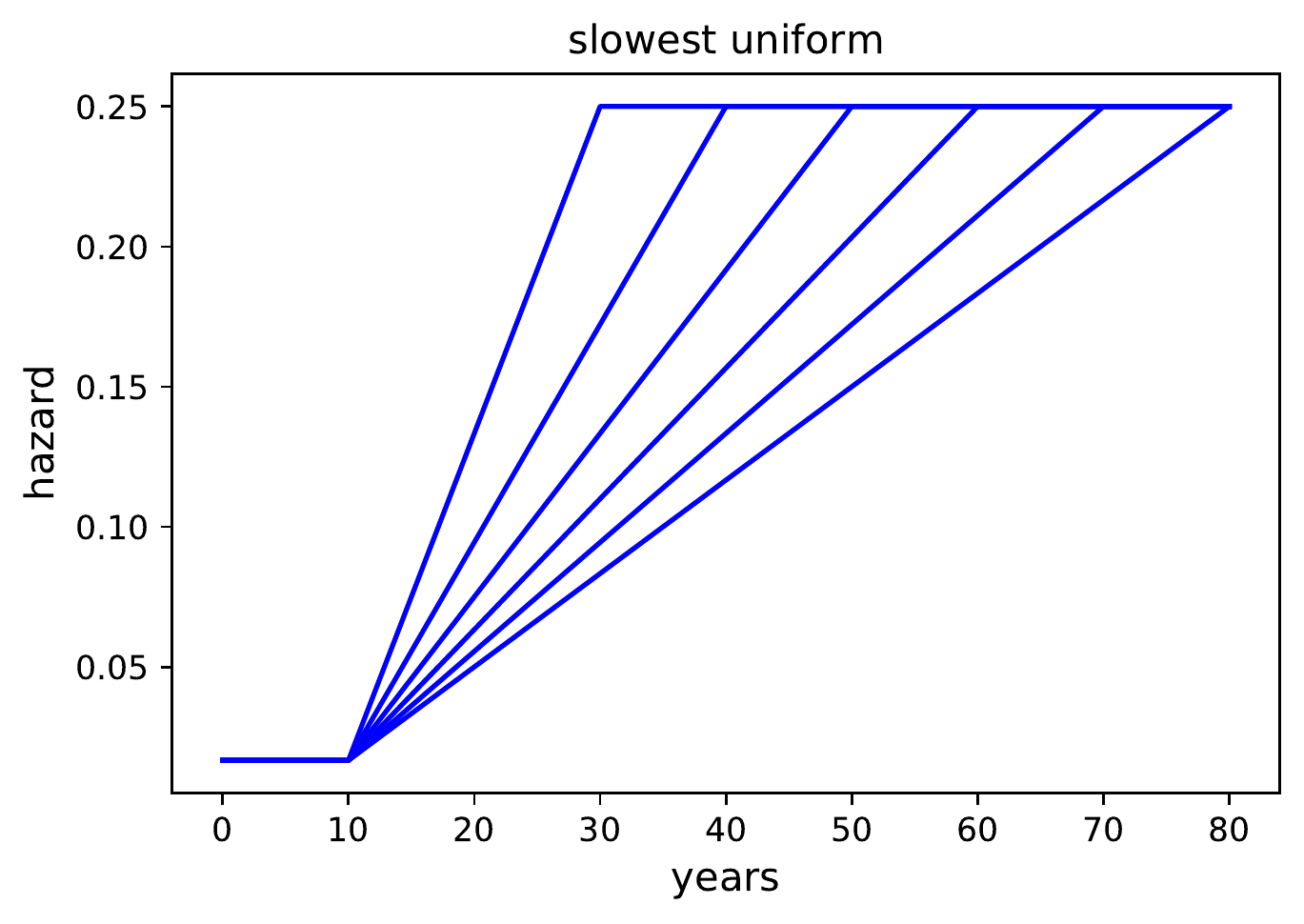}
			\includegraphics[trim=0 0 0 0, clip, width=0.65\textwidth]{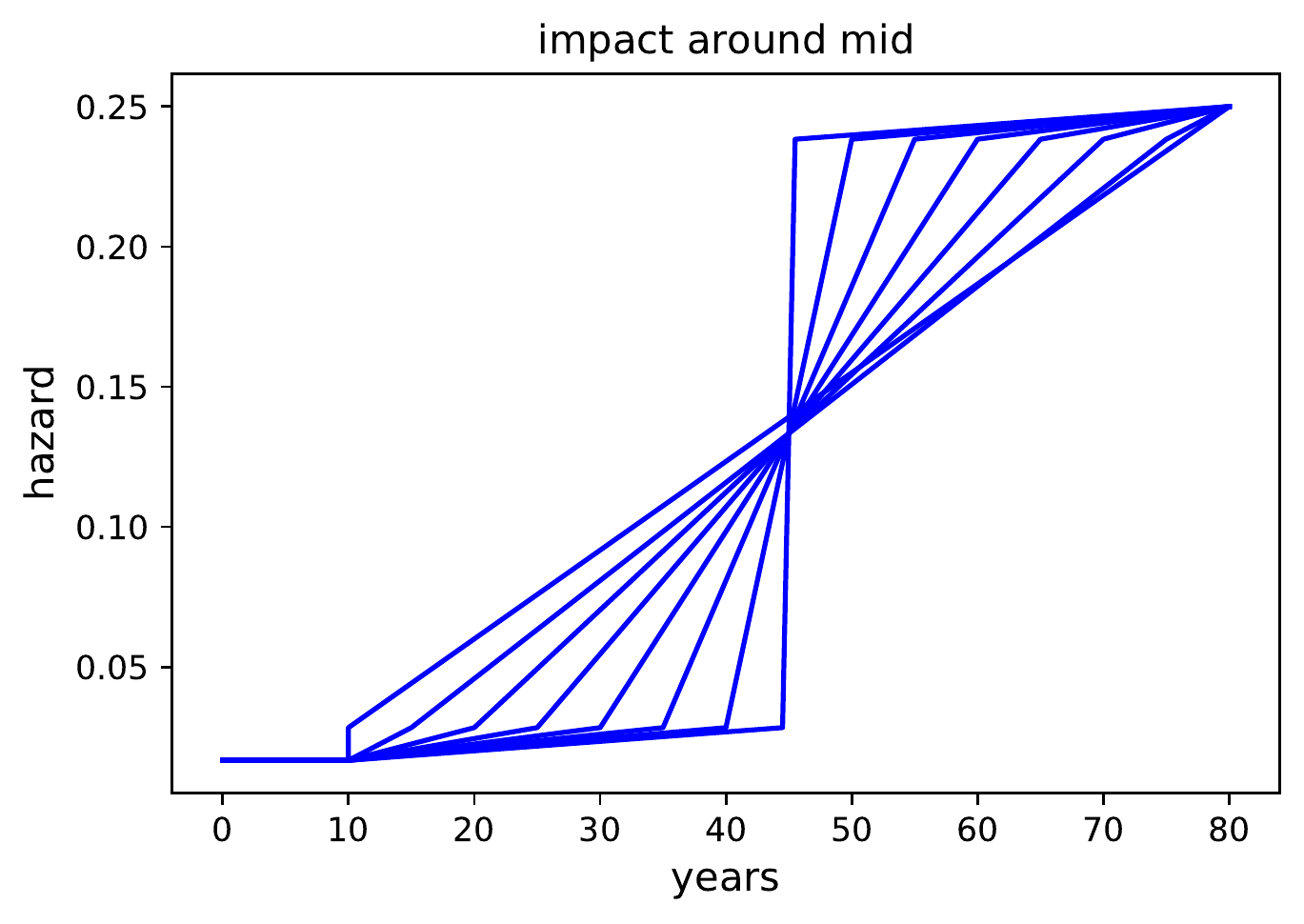}
		\end{center}
		\caption{Slowest uniform test case approaches of instantaneous hazard rate to climate change endpoint, starting from CDS of 100bps up to 10 years  on LEFT above.   On RIGHT test cases when the impact is around the mid point from now to 2101.  \XVA\ impacts are given in Sections \ref{s:slow} and \ref{s:mid}.}
		\label{f:comp}
	\end{adjustwidth}
\end{figure}

Figure \ref{f:comp} compares plots of the instantaneous hazard rates.  Note that because $u=5\%$, i.e. there is a build-up, there is also a jump in instantaneous hazard rate at the switch from \Qbb\ to \Pbb\ for the slowest increase.

\begin{table}
	\begin{adjustwidth}{-0cm}{-0cm}
		\begin{center}
			\begin{tabular}{lrrrr}
				\toprule
				{} & \multicolumn{4}{l}{change in CVA \%, 30Y IRS} \\
				IRS length (years) &  \qquad       20 &   \qquad   30 &  \qquad    40 &   \quad   50 \\  \cline{1-1}
				width (years) &            &        &        &        \\
				\midrule
1     &        2 &   8 &  10 &  16 \\
10    &        3 &   9 &  11 &  18 \\
20    &        3 &  10 &  14 &  24 \\
30    &        4 &  13 &  19 &  31 \\
40    &        6 &  19 &  29 &  40 \\
50    &        8 &  32 &  44 &  53 \\
60    &       18 &  54 &  64 &  69 \\
70    &       42 &  82 &  88 &  89 \\
				\bottomrule
			\end{tabular}
			
			\vskip5mm
			
			\begin{tabular}{lrrrr}
				\toprule
				{} & \multicolumn{4}{l}{change in FVA \%, 30Y IRS} \\
				IRS length (years)&      \qquad    20 &  \qquad   30 &   \qquad  40 &  \quad    50 \\\cline{1-1}
				width (years)&            &       &       &       \\
				\midrule
1     &       -0 &  -1 &  -1 &  -1 \\
10    &       -0 &  -1 &  -1 &  -1 \\
20    &       -0 &  -1 &  -1 &  -2 \\
30    &       -0 &  -1 &  -2 &  -2 \\
40    &       -0 &  -2 &  -2 &  -3 \\
50    &       -0 &  -3 &  -4 &  -5 \\
60    &       -1 &  -6 &  -6 &  -8 \\
70    &       -2 & -10 & -11 & -13 \\
				\bottomrule
			\end{tabular}
			
			\vskip 5mm
			
			\begin{tabular}{lrrrr}
				\toprule
				{} & \multicolumn{4}{l}{change in \XVA\ \%, 30Y IRS} \\
				IRS length (years)&      \qquad    20 &  \qquad   30 &  \qquad   40 & \ \    50 \\\cline{1-1}
				width (years)&            &       &       &       \\
				\midrule
1     &        1 &   4 &   5 &   9 \\
10    &        1 &   4 &   6 &  11 \\
20    &        2 &   5 &   8 &  14 \\
30    &        2 &   6 &  11 &  18 \\
40    &        3 &   9 &  16 &  24 \\
50    &        4 &  16 &  24 &  31 \\
60    &        9 &  26 &  34 &  40 \\
70    &       22 &  39 &  46 &  50 \\
				\bottomrule
			\end{tabular}
			
		\end{center}
		\caption{Impact around mid point to 2101 for instantaneous hazard rate, and $u=05$.  Changes in CVA (top), FVA (mid), and CVA+FVA (bottom), i.e. relative sizes of CD.CVA, CD.FVA, and CCVA compared to flat CDS extrapolation.  Notice that increased hazard rates is slightly beneficial for FVA but not so for CVA. FVA and CVA are different sizes so the overall result is not a simple average.}
		\label{t:exmid}
	\end{adjustwidth}
\end{table}

\FloatBarrier

\subsection{Transition quantification}   \label{s:transQuant}

Table \ref{t:transQ} shows the effect on \XVA\ and survival probabilities within the transition stress $t_\text{mid\ start}$ to $t_\text{mid\ end}$, with $u=0.05$ and the peak hazard rate at 2500bps, for a 30 year IRS.  We consider mid-transition from 15 years in the future to 75 years in the future, and transition durations of 1 to 10 years.  The counterparty has a traded CDS level of 100bps, and we imagine that the counterparty experiences economic stress from changing their business model to adapt to climate change.  We further assume that if they overcome the transition period then they have the same risk level as at the start, i.e. 100bps.

The lowest table in Table \ref{t:transQ} provides the change in survival probability over the transition period, whether this is 1, 5, or 10 years.  This change in probability provides another way to understand the impact of the transition timing and duration relative to the effects on \XVA.

\begin{table}
	\begin{adjustwidth}{-0cm}{-0cm}
		\begin{center}

\begin{tabular}{lrrrrrrr}
\toprule
{} & \multicolumn{7}{l}{change in CVA \%} \\
time to mid  (years)&       15 &  25 &  35 & 45 & 55 & 65 & 75 \\\cline{1-1}
width (years)&            &       &       &      &      &      &      \\
\midrule
1     &       47 &  26 &  10 &  8 &  6 &  5 &  4 \\
5     &      112 &  54 &  11 &  8 &  6 &  5 &  4 \\
10    &      161 &  81 &  13 &  9 &  6 &  5 &  4 \\
\bottomrule
\end{tabular}
			
			\vskip 5mm
			
\begin{tabular}{lrrrrrrr}
\toprule
{} & \multicolumn{7}{l}{change in FVA \%} \\
time to mid (years) &       15 & 25 & 35 & 45 & 55 & 65 & 75 \\\cline{1-1}
width (years) &            &      &      &      &      &      &      \\
\midrule
1     &       -7 & -2 & -1 & -1 & -1 & -0 & -0 \\
5     &      -19 & -4 & -1 & -1 & -1 & -1 & -0 \\
10    &      -29 & -6 & -1 & -1 & -1 & -1 & -0 \\
\bottomrule
\end{tabular}
			
			\vskip 5mm
			
\begin{tabular}{lrrrrrrr}
\toprule
{} & \multicolumn{7}{l}{change in \XVA \%} \\
time to mid (years) &       15 &  25 & 35 & 45 & 55 & 65 & 75 \\\cline{1-1}
width (years) &            &       &      &      &      &      &      \\
\midrule
1     &       22 &  13 &  5 &  4 &  3 &  2 &  2 \\
5     &       52 &  27 &  6 &  4 &  3 &  2 &  2 \\
10    &       73 &  41 &  6 &  4 &  3 &  3 &  2 \\
\bottomrule
\end{tabular}
			
			\vskip 5mm
			
\begin{tabular}{lrrrrrrr}
\toprule
{} & \multicolumn{7}{l}{percent change in survival probability} \\
time to mid  (years)&            15 &  25 &  35 &  45 &  55 &  65 &  75 \\\cline{1-1}
width (years) &                 &       &       &       &       &       &       \\
\midrule
1     &            -9 &  -7 &  -5 &  -4 &  -3 &  -3 &  -2 \\
5     &           -34 & -27 & -21 & -16 & -13 & -10 &  -8 \\
10    &           -51 & -40 & -31 & -24 & -19 & -15 & -12 \\
\bottomrule
\end{tabular}

		\end{center}
		\caption{Impact of transformation stress for 30 year IRS, depending on timing (mid point) and duration (width).  Changes in CVA (top), FVA (mid-upper), and CVA+FVA (mid-lower), and change in default probability over the transformation period (1, 5, or 10 years), i.e. relative sizes of CD.CVA, CD.FVA, and CCVA compared to flat CDS extrapolation. }
		\label{t:transQ}
	\end{adjustwidth}
\end{table}

\FloatBarrier

\section{Discussion}

We introduce \CCVAll\ to capture currently invisible economic impact on credit losses and funding from climate change in as much as this is different to market implied \XVA\ using current CDS extrapolation.  We also provide a rigorous basis both in terms of probability spaces and measures, and in terms of contrasting of potential climate change effects with market practice.

Surprisingly, we find that even for climate change endpoints as far away as 2101, if there is the slowest possible uniform increase of hazard rates, then there are significant credit impacts even on 20y IRS.  We also see that the effect on FVA is opposite in sign to the effect of CVA, simply because increased default probability means less time  paying funding costs.  However, the overall effect is still an increase of \XVA.

Transition effects, unsurprisingly, depend on when they occur and their duration.  Our modeling enables this to be captured with a few clearly interpretable parameters that can then form the basis of discussion with stakeholders, e.g. the risk department, or regulators.

The parameterized approach we introduce for the instantaneous hazard rate curve enables simple comparison and communication of climate change economic impacts whatever the details of the upstream models.

\section{Acknowledgments}

The authors would like to gratefully acknowledge discussions with Robert Wendt, Cathyrn Kelly, Nicholas Amery, Shaoti Zeng, and Astrid Leuba.

\bibliographystyle{chicago}
\bibliography{cchange}

\end{document}